\documentclass[twocolumn,showpacs,amsmath,amssymb,prl,graphics,graphicx]
{revtex4}
      \usepackage{graphicx}
\usepackage{bm}
\usepackage{dcolumn}
\begin{document}

\title{Interaction of vortices in  thin superconducting  films
   and  Berezinskii-Kosterlitz-Thouless transition}

\author{V.G. Kogan,  }
\affiliation{Ames Laboratory - DOE and Department of  Physics and Astronomy,
Iowa State University, Ames IA  50011-3160}

                  \date{\today}
\begin{abstract}
The precondition for the BKT transition in thin superconducting films,
the logarithmic intervortex interaction, is satisfied at distances
short relative to $\Lambda=2\lambda^2/d$, $\lambda$ is the London
penetration depth of the bulk material and $d$ is the film thickness.
For this reason, the search for the transition has  been
conducted in samples of the size $L<\Lambda$. It is argued below that
film edges turn the interaction into near exponential
(short-range) thus making the BKT transition impossible. If however the
substrate is superconducting and separated from the film by an insulated
layer, the logarithmic intervortex interaction  is
recovered and the BKT transition should be observable.
\end{abstract}
\maketitle


   {\it Introduction.} The seminal prediction by Berezinskii
\cite{Ber}, Kosterlitz, and Thouless
\cite{KT} (BKT) of the  phase transition caused by fluctuations-induced
topological excitations in two-dimensional (2D) systems has been confirmed in a
number of experiments with superfluid films. Many attempts to do the same with
thin superconducting films were less convincing, for a review of early
theoretical and experimental work see, e.g., Ref.\,\onlinecite{Minn}.
Later transport data on superconducting films remained inconclusive
\cite{Repaci,Lobb}. The recent transport measurements \cite{TKK} on high
quality ultra-thin high-$T_c$ superconducting films failed to show the jump in
the exponent  $a$ of current-voltage characteristics,
$V\propto I^a$, a signature of the BKT transition \cite{Nelson}.
The discussion persists up to this day \cite{Martinoli,Schneider}.

The precondition for the BKT transition,
the logarithmic intervortex interaction, is satisfied only at distances
short relative to $\Lambda=2\lambda^2/d$ ($\lambda$ is the London
penetration depth of the bulk material and $d$ is the film thickness).
For this reason, the search for the transition has  been
conducted in samples of the size $L<\Lambda$. It is argued here that
even in small samples the interaction is not logarithmic: due to the
boundary conditions at the film edges  the interaction turns into a
short-range near exponential decay. The main conclusion therefore
is that the BKT transition could not happen in thin superconducting
films of any size on insulating substrates (unlike in layered compounds where
the interaction between pancake vortices is logarithmic \cite{***}).
If however the film is placed on a superconducting substrate and
separated from it by an insulating layer of the thickness $s$, the
logarithmic interaction on all distances greater than $s$ is
recovered and the BKT transition should take place.

The situation is different for ac response of thin films. The
characteristic separation of vortices $l_\omega\propto 1/\omega$
contributing to the 2D response might be small at large frequencies
$\omega$ and therefore can be recorded even in small samples,
see Ref.\,\onlinecite{Gasparov} and references therein. We do not
consider ac phenomena here.\\

%
   {\it Approach.} We begin with a brief review of an approach to
    vortices in thin  films suggested in
Ref.\,\cite{Pearlvortex}; although not common this approach allows
one to evaluate energies in a direct manner, the advantage relevant
for our purpose.
           As was stressed by Pearl \cite{Pearl1}, a large contribution to the
vortex energy  in  thin films comes
from  stray fields. In fact, the problem of a vortex in a thin
film is reduced to that of the field  distribution in free space subject
to certain boundary conditions at the film surface. Since
${\rm curl}\,{\bm h}={\rm div}\,{\bm h}=0$ out the film, one can
introduce a {\it scalar} potential for the {\it outside} field:
\begin{equation}
{\bm h}=\nabla \varphi\,,\qquad \nabla^2\varphi =0\,.
\label{eq1}
\end{equation}
To formulate the boundary conditions for the outside Laplace
problem, consider a film of thickness $d\ll \lambda $ occupying the
$xy$ plane. For a vortex at ${\bm r}=0$, the London equations for the
film interior read after averaging over the film thickness:
\begin{equation}
h_z+{2\pi\Lambda\over c}\,{\rm curl}_z {\bm g}=
\phi_0\,\delta ({\bm r} )\,,
\label{eq2}
\end{equation}
where ${\bm g}({\bm r})$ is the sheet current density, ${\bm r}=(x,y)$ and
$\phi_0$ is the flux quantum.

In a thin film, the Maxwell equation
$4\pi{\bm j}/c={\rm curl}\,{\bm h}$  is reduced to relations
between the sheet current and discontinuities of the tangential
fields:
\begin{equation}
               4\pi \bm g /c= \bm {\hat z}\times [\,\bm h (+0)-\bm h (-0)]\,.
\label{eq3}
\end{equation}
One substituts Eq.\,(\ref{eq3}) in  (\ref{eq2}) and uses
div$\,{\bm h}=0$ to obtain:
\begin{equation}
h_z - \Lambda[\,\partial_z h_z(+0) -\partial_z h_z(-0)]/2  =\phi_0
\delta ({\bm r} )\,.
\label{hz}
\end{equation}
This equation expressed in terms of the potential $\varphi$
along with conditions at infinity constitute the boundary
conditions for the Laplace problem, Eq.\,(\ref{eq1}), for the field
distribution outside the film.

Consider the case in which the  half-spaces above and
under the film are vacuum (or a
non-magnetic insulator).
                The general form of the potential that vanishes at $z\rightarrow
+\infty$ of the empty upper half-space is
\begin{equation}
\varphi ({\bm r},z)=\int\frac{d^2{\bm k}}{(2\pi)^2}\,
\varphi({\bm k})\,e^{i{\bm k}\cdot{\bm r}-kz}\,.
\label{phi_general}
\end{equation}
Here, $\varphi({\bm k})$ is the 2D Fourier transform and
$k^2= k_x^2+k_y^2 $. In the lower half-space one has to replace
$z\rightarrow -z$. Due to the symmetry, the boundary condition
(\ref{hz}) becomes $\partial_z\varphi(+0)- \Lambda \partial_z^2
\varphi(+0)=\phi_0\delta ({\bm r} )$ that yields the Fourier
transform of the potential in the upper half-space:
\begin{equation}
\varphi  =-\frac{\phi_0}{k(1+ k\Lambda)}\,.
\label{eq6}
\end{equation}
This provides  $h_z({\bm k})=-k\varphi({\bm k})$ and
$h_{x,y}({\bm k})=ik_{x,y}\,\varphi({\bm k})$, i.e. the fields
everywhere in the free space as well as the sheet currents
     due to a single vortex in an {\it infinite}  film. One
can easily verify that this solution coincides with that
given by Pearl \cite{Pearl1}.\\

   {\it Energy.} We now establish connection between vortex energy
and the potential $\varphi$. We begin with
the general situation of a vortex in a {\it finite} bulk sample.
The energy consists of the London energy (magnetic + kinetic) inside the
sample, $\epsilon^{(i)}=\int [h^2+(4\pi\lambda j/c)^2]dV/8\pi$ and the
magnetic energy outside, $\epsilon^{(a)}=\int h^2dV/8\pi$.
Then, for the potential introduced in Eq.\,(\ref{eq1}) and
gauged to zero at $\infty$ (which is possible in {\it zero applied field})
one has $8\pi \epsilon^{(a)}=\oint\varphi{\bm h}\cdot d{\bm S}\,$
where the integral is over the sample surface with $d{\bm S}$ directed
inward the material. The London part is transformed integrating by parts the
kinetic term: $8\pi \epsilon^{(i)}=(4\pi\lambda^2/c)\oint({\bm j}\times {\bm
h})\cdot d{\bm S}$ where the integral is over the sample surface and
the surface of the vortex core. The integral over the sample surface
is further transformed:
$\oint d{\bm S}\cdot ({\bm j}\times \nabla\varphi)=\oint d{\bm S}\cdot
\varphi (\nabla \times {\bm j})$  \cite{rem1}.
                 Combining the result with $\epsilon^{(a)}$, one
obtains $\oint d{\bm S}\cdot \varphi ({\bm h}+4\pi\lambda ^2{\rm
curl}{\bm j}/c)$ where the expression in parenthesis is $\phi_0 \hat{{\bm
v}}\delta^{(2)}({\bm r}-{\bm r}_v)$ where
$\hat{{\bm v}}$ is the direction of the vortex crossing the surface at the
point ${\bm r}_v$, and $\delta^{(2)}({\bm r}-{\bm r}_v)$ is the 2D
$\delta$ function. One then obtains:
\begin{equation}
\epsilon=\frac{\phi_0}{8\pi}[\varphi({\bm r}_{ent})-\varphi({\bm
r}_{ex})]-
\frac{ \lambda ^2}{2c}\oint_{core} d{\bm S}\cdot ({\bm h}\times {\bm
j})\,;
\label{energ}
\end{equation}
    ${\bm r}_{ent}$ and ${\bm r}_{ex}$ are the positions of the
vortex entry and exit at the sample surface (the vortex is assumed
to cross the sample surface at right angles; otherwise, one should
multiply the potentials by cosines of corresponding angles). For
more than one vortex one has to sum up expressions (\ref{energ})
over all vortices. If more than one superconductor is present,
Eq.\,(\ref{energ}) still holds, but $\varphi$ as a solution
of Laplace equation outside superconductors will be affected by
the presence of all superconductors.

For thin films, the integral over the core surface ($\propto d$) can
be neglected:
\begin{equation}
\epsilon=\frac{\phi_0}{8\pi}\sum_\nu D\varphi_\nu
\label{energ_films}
\end{equation}
where the notation $D\varphi_\nu\equiv  \varphi_\nu({\bm
r}_{ent})-\varphi_\nu({\bm r}_{ex}) $ for the $\nu$-vortex is
introduced for brevity.
Due to linearity of the Laplace and London equations, one has for
two vortices
$\varphi=\varphi_1+\varphi_2$ and
              \begin{eqnarray}
\epsilon &=&\frac{\phi_0}{8\pi}\,[D\varphi (1)+D\varphi
(2)] \label{two}\\
&=&\frac{\phi_0}{8\pi}D[\varphi_1(1)+ \varphi_2(1)+
    \varphi_2(1)+ \varphi_2(2)]
\nonumber\\
\nonumber
\end{eqnarray}
where the arguments 1,2 are positions of vortices.
Clearly, the self-energy of the first vortex is
              \begin{eqnarray}
              \epsilon^{(0)}_1 =\frac{\phi_0}{8\pi}\,D\varphi_1(1)
\label{e_self}
\end{eqnarray}
and the interaction energy is given by
              \begin{eqnarray}
\epsilon_{int} = \frac{\phi_0}{8\pi}\,D[ \varphi_2(1)+ \varphi_2(1)]\,.
\label{e_int0}
\end{eqnarray}


   {\it Infinite film in vacuum.}   Equations\,(\ref{e_self})
and (\ref{e_int0}) are quite general and hold for films of any
lateral size. In particular, for an {\it infinite} film in vacuum
one obtains with the help of the potential (\ref{eq6}):
              \begin{eqnarray}
              \epsilon^{(0)} =\frac{\phi_0}{4\pi}\int\frac{d^2{\bm
k}}{(2\pi)^2}\,\frac{\phi_0}{k(1+k\Lambda)}
={\phi_0^2\over 8\pi^2\Lambda}\ln {\Lambda\over\xi}\,,
\label{self}
\end{eqnarray}
where the cutoff at $k_{\rm max}\approx 1/\xi$ is introduced to a
logarithmically divergent integral. Similarly, one finds:
              \begin{eqnarray}
              \epsilon_{int} =\frac{\phi_0^2}{8\pi\Lambda}\,\left[{\bm
H}_0\Big(\frac{r}{\Lambda}\Big)-Y_0\Big(\frac{r}{\Lambda}\Big)
\right]\,,
\label{e_int1}
\end{eqnarray}
where ${\bm H}_0$ and $Y_0$ are Struve and Bessel functions and $r$
is the distance between vortices. At  distances $r\gg\Lambda$ this
yields $\epsilon_{int}=\phi_0^2/4\pi^2r$ (as  for two point
``magnetic charges" $\phi_0 /2\pi$) showing
that at large distances the role of stray fields is dominant. For
$r\ll\Lambda$, the interaction is logarithmic,
              \begin{eqnarray}
              \epsilon_{int} =\frac{\phi_0^2}{4\pi^2\Lambda}\,\ln
             \frac{2\Lambda}{r}\,,
\label{e_int2}
\end{eqnarray}
        that led to a statement that the BKT transition should be
searched for in samples of the size $L\ll\Lambda$. We show below
that this statement is incorrect.\\


   {\it Small samples in vacuum.} To consider small samples, one
turns back to the basic Eq.\,(\ref{eq2}). The currents ${\bm
g}({\bm r})$ can be found by solving Eq.\,(\ref{eq2}) combined with
the continuity equation and the Biot-Savart integral which relates
the field $h_z$ to the surface current:
\begin{equation}
{\rm div} {\bm g}=0\,,\,\,\,\,\,
h_z({\bm r}) =\int d^2 {\bm r}'[{\bm g}({\bm r}')\times {\bm
R}/cR^3]_z\,;
\label{e3}
\end{equation}
${\bm R}={\bm r}-{\bm r}'$.
To satisfy ${\rm div} {\bm g}=0$ it is convenient to deal with a scalar
stream function
$G({\bm r})$ such that ${\bm g}={\rm curl}G{\hat {\bm z}}$.
Alternatively, the sheet current can be expressed in terms of
the derivatives of the potential $\varphi({\bm r},+0)$ at the upper
film face:
\begin{equation}
                 g_x=c\,\partial_y\varphi/2\pi \,,\quad
g_y=c\,\partial_x\varphi/2\pi\,.
\label{eq15}
\end{equation}
Therefore, $G({\bm r})$ is proportional to $\varphi({\bm r},z=+0)$:
\begin{equation}
\varphi({\bm r},+0)=-2\pi G({\bm r})/c
\label{G-phi}
\end{equation}
(a possible additive constant is zero since both $\varphi$
and $G$ are gauged to zero at $\infty$). Since  the
energies are expressed in terms of  $\varphi$ at
vortex positions, Eq.\,(\ref{G-phi}) shows that to evaluate vortex energies and
their interaction it suffices to find the
current distribution $G({\bm r})$.

The problem of current distribution  for samples  of
arbitrary size is difficult:   the London
Eq.\,(\ref{hz})  combined with Biot-Savart law (\ref{e3}) make an
    integro-differential equation for  $\bm g$. However, for small samples,
$L\ll\Lambda$, it is manageable because as is seen from Eq.\,(\ref{e3}) the
field
$h_z\sim g/c$ whereas the term with current derivatives in
Eq.\,(\ref{eq2}) is of the order $\Lambda g/cL$. Then,
Eq.\,(\ref{eq2}) can be rewritten as a Poisson equation for $G$:
\begin{equation}
\nabla^2G=-(c\phi_0/2\pi\Lambda)\delta({\bm r}-{\bm a})\,.
\label{30}
\end{equation}

The normal to the edge component of the current vanishes, i.e.,
$G$ is a constant along the edges which can be set zero. Thus, $G$
is equivalent to  the electrostatic potential of a linear charge
$q=c\phi_0/8\pi^2\Lambda$ situated at the vortex position ${\bm a}$
and parallel to the side surface of the grounded metal cylinder with the
crossection coinciding with the thin-film sample.  A rich
library of the 2D electrostatics is instrumental in solving for vortex
currents for a variety of film shapes with linear dimensions less than
$\Lambda$; see, e.g., Ref.\,\cite{Morse} for the solution of Eq.\,(\ref{30})
for a rectangular film.\\

   {\it Thin-film strips in vacuum.} Most of  experiments of our
interest are done on long thin-film strips. For a strip along $y$
with edges at $x=0$ and $x=W\ll\Lambda$, the solution of Eq.\,(\ref{30}) for a
vortex placed at ${\bm a}=(a_x,a_y)$ with zero boundary conditions at the
edges reads:
\FL
\begin{equation}
{\rm tanh}\frac{G}{2q} = \frac{{\rm
sin}(\pi a_x)\, \,{\rm sin}(\pi x)}{{\rm cosh}[\pi (y-a_y)]-
{\rm cos}(\pi a_x)\,{\rm cos}(\pi x)}\,,
\label{41}
\end{equation}
where for brevity $W$ is used as a unit length (see, e.g.,
Refs.\,\cite{Morse}) or \cite{Smithe}. Due to the edges,
        $G$ does not depend exclusively on $|{\bm r}-{\bm a}|$; note however
that $G({\bm r},{\bm a})=G({\bm a},{\bm r})$. It is
straightforward now  to find the energies of vortices in a strip and their
interaction with the help of general formulas
             (\ref{e_self}), (\ref{e_int0}),  and (\ref{G-phi}).

The self-energy of a vortex at the position $x=a$, $y=0$,
$\epsilon=\phi_0G(a)/2c$, is logarithmically divergent.
Introducing the cutoff at $|{\bm r}-{\bm a}|=\xi$ one obtains
\cite{Pearlvortex}:
\begin{equation}
\epsilon^{(0)}(a)=\frac{\phi_0^2}{8\pi^2\Lambda}\,{\rm ln}\left(
\frac{2W}{\pi\xi}\,{\rm sin}\frac{\pi a}{W}\right )\,.
\label{vortex-strip}
\end{equation}

The interaction  (\ref{e_int0}) of two vortices at
${\bm a}_1$ and ${\bm a}_2$ is
\begin{equation}
\epsilon_{int}= \frac{\phi_0^2}{8\pi^2\Lambda} \ln\frac{
\cosh[\pi (y_1-y_2)]-\cos[\pi (x_1+x_2)]}
{ \cosh[\pi (y_1-y_2)]-\cos[\pi (x_1-x_2)]}
\label{41}
\end{equation}
where the coordinates are given again in units of $W$. (Formally similar
interaction exists between Abrikosov vortices parallel to the
thin film \cite{Gurevich}.)
Hence the interaction is not proportional to the logarithm of the
intervortex distance as required by BKT (except at distances of the
order $\xi$ where the present theory breaks down). E.g., for $y_1=y_2$ we
have (in conventional units):
\begin{equation}
\epsilon_{int}= \frac{\phi_0^2}{4\pi^2\Lambda} \ln\Big|\frac{
            \sin[\pi (x_1+x_2)/2W]}{\sin[\pi (x_1-x_2)/2W]}\Big|\,.
\label{41}
\end{equation}
For vortices in the strip middle ($x_1=x_2=W/2$) separated by $y$,
\begin{equation}
\epsilon_{int}\approx\frac{\phi_0^2}{4\pi^2\Lambda} \ln\Big|\coth\left(
            \frac{ \pi y}{2W}\right)\Big|\,.
\label{42}
\end{equation}
This gives for $y>W$:
\begin{equation}
\epsilon_{int}= \frac{\phi_0^2}{2\pi^2\Lambda} \exp\left(-\,\frac{
            2\pi y}{W}\right)\,.
\label{41}
\end{equation}
            Thus, for separations larger than $W/2\pi$, the intervortex
interaction is in fact exponentially weak and
short-range. Physically, this happens because the second vortex
feels not only the ``bare" first, but also the infinite chain of
$\pm$ images which one has to introduce to satisfy the boundary
conditions at the edges.
The possibility of BKT transition in thin-film bridges of YBCO
similar to those studied in \cite{TKK} has been examined
experimentally \cite{Repaci} and theoretically \cite{Lobb} (see
      references therein) with no positive outcome. Given the above
argument, one can say that the short range interaction in small
samples is the reason for this failure.\\


   {\it Superconducting substrate.} Up to this point, we have
discussed a ``free standing" film or, better to say, a thin
superconducting film placed on an insulating non-magnetic substrate
so that the symmetry of the upper and lower half-spaces could be
utilised. The situation changes if the film is placed on a
superconducting substrate being separated from it by an insulating
layer to prevent the Josephson coupling to the substrate. The
vortex  magnetic flux
$\phi_0$ is channeled into the space between
the film and the screening substrate so that the radial field component
along with azimuthal sheet currents vary as $1/r$ at large distances. This
results in the  Lorentz interaction force of two vortices varying as $1/r$
and, therefore, in the logarithmic interaction energy.

To confirm this quantitatively, consider a film situated at a
   $z=s$ above a superconducting substrate occupying the half-space
$z<0$. If $s$ is large enough to suppress the Josephson interaction, the
problem is reduced to solving the Laplace equation (\ref{eq1}) for the
potential $\varphi_A({\bm r},z)$ in the free space $z>s$ (the domain A)
and for $\varphi_B({\bm r},z)$ in the domain B between the film
and the substrate,  $0<z<s$. A simple method of doing this is described in
Ref.\,\onlinecite{MKC}. One looks for
$\varphi_A({\bm r},z)$ in the form (\ref{phi_general}) with the 2D Fourier
transform $\varphi_A({\bm k})\,e^{-kz}$. Similarly, in the finite domain
B, $\varphi_B({\bm k},z)=C_{1}({\bm k} )\,e^{kz}+C_{2}({\bm
k} )\,e^{-kz}$. The boundary conditions at $z=s$ are given by the field
continuity and by the London equation (\ref{eq2}).
One further simplifies the  problem of the substrate screening by
setting  the substrate penetration depth to zero, i.e., $h_z(z=0)=0$
at the substrate surface; this gives $C_{1}=C_{2}\equiv C$. After
simple algebra one obtains:
\begin{equation}
\varphi_A=-2 C \,e^{ks}\sinh\,ks
=-\frac{\phi_0\,e^{ks}\sinh\,ks}{k(\sinh\,ks+k\Lambda\,e^{ks}/2)}.
\label{phis}
\end{equation}

Given the potentials above and under the film, one uses
Eq.\,(\ref{e_self}) to calculate the   vortex energy:
\begin{eqnarray}
\epsilon
=\frac{\phi_0^2}{ 16\pi^2}\int\frac{dk\,e^{ks} }{\sinh\,ks
       +k\Lambda\,e^{ks}/2}\,,
\label{single_energ}
\end{eqnarray}
The lower limit of integration is the inverse sample size $1/L$, whereas the
upper one is $1/\xi$. One can estimate this integral splitting the
integration domain in two: $s/\Lambda<ks<1$ and $1<ks<s/\xi$. The
contributions to the integral are estimated as $(2/\Lambda)\ln(L/s)$ and
$(4/\Lambda)\ln(s/\xi)$. This gives
\begin{eqnarray}
\epsilon\approx\frac{\phi_0}{4\pi^2\Lambda} \ln\frac{\sqrt{Ls}}{\xi}\,.
\label{single_energ1}
\end{eqnarray}

The interaction energy  is found with the help of the general result
(\ref{e_int0}):
    \begin{eqnarray}
    \epsilon_{int} =
\frac{\phi_0^2}{8\pi^2}\int_{1/L}^{1/\xi}
dk\,J_0(kr)\frac{\cosh\,ks+e^{ks}\sinh\,ks}{\sinh\,ks
       +k\Lambda\,e^{ks}/2}.
\label{e_int3}
\end{eqnarray}
The oscillating Bessel function truncates the integral at $k\approx
1/r$, so that in the relevant part of the integration
domain  $ks < s/r\ll 1$ for intervortex distances exceeding $s$. One
readily estimates:
              \begin{eqnarray}
              \epsilon_{int} \approx\frac{\phi_0^2}{4\pi^2\Lambda}\,\ln
             \frac{ \Lambda}{r}\,,\qquad r > s\,.
\label{e_int4}
\end{eqnarray}
       Thus, the superconducting film situated parallel to the
surface of a bulk superconductor should exhibit the BKT transition, unlike the
case of insulating substrate, a verifiable conclusion. \\

To summarize, it is shown that in small thin-film samples on insulating
substrates, edge effects modify the vortex-vortex interaction making it
short-range, unlike the logarithmic long-range interaction
needed for the BKT transition. In large free-standing films this
transition cannot happen because of the $1/r$ interaction via the
stray fields. This makes the BKT transition impossible in thin
films of any size if they are supported by a non-superconducting
substrate. On the contrary, if the substrate is superconducting and
separated from the film by an insulated layer of a thicknes  $s$,
the interaction is logarithmic at all distances $r>s$ and
the BKT transition should be observable in a dc type of experiment;
e.g., the dc power-law current-voltage characteristics should show
the well-known jump of the power exponent at the BKT transition.

The idea of this work emerged while studying the IV characteristics of
thin-film bridges taken by F. Tafuri and J. Kirtley. Also, the author is
indebted to L. Bulaevskii and R. Mints for  helpful discussions. The work
is supported by the Office of Basic Energy Sciences, US DOE. Ames
Laboratory is operated for DOE by the Iowa State University under
Contract No. W-7405-Eng-82.

\end{document}